**Multi-modal Spectroscopic Study of Surface Termination Evolution in $Cr_2TiC_2T_x$ MXene**

*James L. Hart, Kanit Hantanasirisakul, Andrew C. Lang, Yuanyuan Li, Faisal Mehmood, Ruth Pachter, Anatoly I. Frenkel, Yury Gogotsi, Mitra L. Taheri\**

Dr. J. L. Hart,[+] Prof. M. L. Taheri
Department of Materials Science and Engineering, Johns Hopkins University, Baltimore, MD, 21218, USA
E-mail: mtaheri4@jhu.edu

K. Hantanasirisakul, Prof. Y. Gogotsi
Department of Materials Science and Engineering, Drexel University, Philadelphia, PA 19104, USA
A. J. Drexel Nanomaterials Institute, Drexel University, Philadelphia, PA 19104, USA

Dr. A. C. Lang
American Society for Engineering Education Postdoctoral Fellow, U.S. Naval Research Laboratory, Washington, DC 20375, USA

Dr. Y. Li, Prof. A. I. Frenkel
Department of Materials Science and Chemical Engineering, Stony Brook University, Stony Brook, NY, 11790, USA

Prof. A. I. Frenkel
Division of Chemistry, Brookhaven National Laboratory, Upton, NY, 11973, USA

Dr. F. Mehmood, Dr. R. Pachter
Air Force Research Laboratory, Materials and Manufacturing Directorate, Wright-Patterson Air Force Base, OH, 45433, USA

[+] present address: Department of Mechanical Engineering and Materials Science, Yale University, New Haven, CT 06511, USA

Control of surface functionalization of MXenes holds great potential, and in particular, may lead to tuning of magnetic and electronic order in the recently reported magnetic $Cr_2TiC_2T_x$. Here, vacuum annealing experiments of $Cr_2TiC_2T_x$ are reported with in situ electron energy loss spectroscopy and novel in situ Cr $K$-edge extended energy loss fine structure analysis, which directly tracks the evolution of the MXene surface coordination environment. These in situ probes are accompanied by benchmarking synchrotron X-ray absorption fine structure measurements and density functional theory calculations. With the etching method used here, the MXene has an initial termination chemistry of $Cr_2TiC_2O_{1.3}F_{0.8}$. Annealing to 600 °C results in the complete loss of –F, but =O termination is thermally stable up to (at least) 700 °C. These findings demonstrate thermal control of –F termination in $Cr_2TiC_2T_x$ and offer a first step towards termination engineering this MXene for magnetic applications. Moreover, this work demonstrates high energy electron spectroscopy as a powerful approach for surface characterization in 2D materials.

**1. Introduction**

MXenes are a large family of 2D materials which show exceptional performance in applications such as energy storage, electromagnetic interference shielding, catalysis, and wireless electronics.[1–8] Their general formula is $M_{n+1}X_nT_x$, where M is a transition metal, X is C and/or N, $n = 1 – 4$, and T represents the surface termination. With typical solution-processed synthesis methods, T is a non-uniform mixture of –OH, –F and =O.[9] Modification of the MXene surface termination holds great potential for engineering functional properties. For instance, experiments have demonstrated that altering T can tune the MXene's metallic conductivity, work function, charge storage properties, and superconductivity.[10–14] Despite a few notable exceptions,[10] termination-property coupling studies have largely focused on $Ti_3C_2T_x$, and thus there is a need to explore the rest of the MXene family. Additionally, improved methods must be developed for effective control of the MXene termination. A primary challenge is removal of the initially mixed surface functionalization without damaging the underlying $M_{n+1}X_n$ material. For example, during vacuum annealing of $Ti_3C_2T_x$, secondary Ti oxides form prior to complete –F removal and before any loss of =O.[14] Another challenge is reliably tracking termination changes, as different techniques tend to report different termination concentrations.[14–18] Ideally, methods capable of directly measuring the surface termination in situ should be developed.

A particularly interesting MXene with regards to surface functionalization is $Cr_2TiC_2T_x$ (shown in **Figure 1**). Recently, we showed that this MXene undergoes a magnetic transition near ~30 K, adding an exciting new functionality to the MXene family.[19] However, as opposed to ferromagnetic or antiferromagnetic order, no long-range magnetic order was experimentally detected. This may be related to the non-uniform mixture of –OH, –F and =O surface terminations, which according to density functional theory (DFT) predictions, favor different magnetic structures.[20,21] DFT has also predicted $Cr_2TiC_2T_x$ to be a rare instance of a semiconducting MXene,[22] with surface terminations controlling the bandgap.[20,21] Thus, understanding and modifying the $Cr_2TiC_2T_x$ surface termination has significant implications for the advancement of magnetic and semiconducting MXenes.

Here, we present an in situ multi-modal spectroscopy study of surface termination evolution in $Cr_2TiC_2T_x$. While annealing $Cr_2TiC_2T_x$ up to 800 °C under vacuum within a transmission electron microscope (TEM), we tracked changes in the O and F atomic concentrations with electron energy loss spectroscopy (EELS). Additionally, by leveraging direct detection EELS at high energy loss (6 keV),[23,24] we measured the Cr $K$-edge and performed extended energy loss fine structure (EXELFS) analysis, which directly probes the MXene surface coordination chemistry. Our EXELFS measurements are in good agreement with DFT calculations and benchmarking synchrotron X-ray absorption fine structure spectroscopy measurements. Together, the composition (via EELS) and coordination environment (via EXELFS) measurements indicate an initial composition of $Cr_2TiC_2O_{1.3}F_{0.8}$. Vacuum annealing at 600 °C leads to the complete loss of –F, but no loss of =O is observed. This data provides an important first step in termination engineering $Cr_2TiC_2T_x$ and presents in situ high energy EELS / EXELFS as a powerful means to study surface chemistry in 2D materials.

## 2. Results and Discussion
Details of the $Cr_2TiC_2T_x$ MXene synthesis, as well as structural and chemical characterization to confirm phase pure MXene, are described in our prior work.[19] Breifly, the parent $Cr_2TiAlC_2$ MAX phase was etched in an aqueous HF/HCl solution, followed by washing in deionized water, and then delamination via tetramethylammonium hydroxide intercalation (see Methods). MXene films were then deposited onto in situ annealing holders for TEM and EELS analysis (see Methods). The studied MXene films consisted of many overlapping flakes with total film thicknesses on the order of 50 – 100 nm (Figure S1). The sample structure was measured through electron diffraction (which is an effective method to detect secondary oxide nanoparticles in MXenes),[14] and initially, only peaks corresponding to MXene were observed (Figure S1). No secondary phases were observed after 700 °C annealing within the TEM vacuum (~$10^{-7}$ Torr); however, after annealing at 800 °C, an additional peak was observed, indicating the onset of MXene decomposition (Figure S1). Thus, for spectroscopic analysis of the MXene surface chemistry evolution, we only consider annealing treatments performed at temperatures ≤ 700 °C.

### 2.1. EELS Composition Analysis
To understand changes in composition with annealing, we measured the Cr $L$-edge, Ti $L$-edge, O $K$-edge, and F $K$-edge in the as-prepared (AP) state and after successive in situ vacuum annealing treatments (**Figure 2a-c**, Figure S2). Only subtle changes are observed in the Cr and Ti $L$-edge fine structure with annealing, which reflects the stability of the MXene structure under the chosen annealing conditions, consistent with the electron diffraction.

The most dramatic spectral change with annealing is observed in the F $K$-edge, where peak $i$ completely disappears after annealing at 600 °C, and the overall edge intensity is greatly reduced (Figure 2c). The overall decrease in the F $K$-edge intensity directly reflects a loss of fluorine from the sample. Regarding the F $K$-edge fine structure, peak $i$ has been attributed to hybridization between –F and MXene surface transition metal $d$ orbitals.[25] The lack of peak $i$ after annealing indicates that no –F terminations remain. Instead, we suggest the residual F $K$-edge intensity (peak $ii$) is due to aluminum fluoride byproduct. Though not detected with electron diffraction, aluminum fluorides can form as byproducts during MXene etching,[26] and the measured F $K$-edge fine structure after 700 °C annealing is consistent with F bonded to Al. The complete removal of –F from $Cr_2TiC_2T_x$ without any detectable secondary oxide formation is an important result; for the widely studied $Ti_3C_2T_x$ and other Ti-based MXenes, –F remains even after annealing at > 700 °C, at which point Ti oxides begin to form. This, in principle, allows the formation of $Cr_2TiC_2O_2$ with homogeneous oxygen termination, which could have a ferromagnetic ground state according to a DFT prediction.[21]

In contrast to the significant decrease in the F $K$-edge intensity, we observed a slight increase in the O $K$-edge intensity with annealing (Figure 2a). Note that the measured O $K$-edge has contributions from several possible sources: =O termination, –OH termination, and $H_2O$ intercalation. Due to the constant (and in fact increasing) O $K$-edge intensity with thermal annealing, we can rule out any sizeable contribution from $H_2O$ intercalation, since $H_2O$ de-intercalation occurs below 300 °C.[14,27] Any loss of $H_2O$ would result in a decrease in the measured O $K$-edge intensity, which is not observed experimentally. Regarding –OH termination, DFT calculations have predicted that for $Cr_2TiC_2T_x$ –OH termination is less stable than –F,[20] and from our F $K$-edge measurements, –F is only stable up to ~500 °C (Figure 2c). Moreover, in Ti-based MXenes, –OH termination is only stable up to ~350 °C.[14,28,29] Hence, the thermal stability of the O $K$-edge spectra is inconsistent with –OH termination in the AP MXene, which would likely desorb within the temperature range of our measurements. Conversely, DFT predicts that =O is the most stable termination for $Cr_2TiC_2T_x$ (compared to –OH and –F),[20] and for Ti-based MXenes, =O is stable well above 700 °C.[14] This reasoning suggests that the measured O $K$-edge spectra are due to =O termination for all annealing conditions. However, it has been reported that the electron beam can transform –OH termination into =O.[30,31] Thus, it is possible that the AP MXene contains some amount of –OH termination which is transformed to =O termination under the electron beam, avoiding the loss of –OH with subsequent annealing.

To explain the slight increase in the O $K$-edge intensity with annealing, we consider two possibilities. First, owing to the presence of non-terminated Cr sites after –F removal, O present within the TEM vacuum could terminate the exposed Cr surface sites, leading to an increase in =O termination. Alternatively, with annealing, the bonding between Cr and =O could change, for instance, due to =O occupying different surface sites after the removal of –F[16] or due to the electron beam transforming –OH into =O. Changes in the hybridization between O and Cr could increase the intensity of certain spectral features in the near edge region of the O $K$-edge spectra. These two scenarios (increase in O content or change in bonding) could be differentiated through measuring the intensity of the O $K$-edge in the post-edge region, but overlap between the Cr $L$-edge and O $K$-edge complicates such analysis. Nevertheless, it is clear that even if there is a slight increase in =O termination, the total amount of surface termination (including both =O and –F) decreases significantly after annealing up to 700 °C.

### 2.2. EXELFS Coordination Analysis

To more directly probe the MXene surface chemistry, we performed a second set of annealing experiments with in situ Cr $K$-edge EELS measurements and EXELFS analysis. The EXELFS signal is sensitive to the radial distribution function of Cr, and therefore contains direct information on the termination of the MXene surface. EXELFS analysis on the Cr $L$-edge is not feasible owing to overlap between the Cr $L_3$, Cr $L_2$, O $K$, and F $K$-edges, which necessitates measurement of the Cr $K$-edge. However, conventional EELS measurements are limited to energies ≤ 2 keV, and spectral analysis at the Cr $K$-edge (~6 keV) has only recently become practical through instrumental innovations.[23,24,32] So, to benchmark our measurements, we performed comparative synchrotron X-ray absorption fine structure (XAFS) measurements of $Cr_2TiC_2T_x$ in the AP state. **Figure 3a,b** present the normalized Cr $K$-edge EELS data and the $R$- space EXELFS signal, respectively. For comparison, we also plot the Cr *K*-edge XAFS data and the extended X-ray absorption fine structure (EXAFS) signal. Despite the fact that the XAFS measurements benefit from improved energy resolution relative to EELS, the data of the two techniques are in good agreement for the AP MXene. We also note that in comparison to XAFS measurements of Cr standards (Cr foil with $Cr^0$, $Cr_2O_3$ with $Cr^{3+}$, and $CrO_3$ with $Cr^{6+}$), the energy onset of the AP MXene data indicates that Cr is in a ~$Cr^{3+}$ state (Figure 3a inset).[33] This finding is consistent with prior XPS data of $Cr_2TiC_2T_x$.[19]

The phase shift corrected, Fourier transform magnitudes of $k^2$-weighted spectra, $|\chi(R)|$, show two main peaks, one at ~1.9 Å and a second at ~2.8 Å (Figure 3b). Note that peaks in $|\chi(R)|$ are related to peaks in the partial radial distribution function of surface Cr atoms. To guide EXELFS interpretation, DFT calculations were performed using various levels of DFT theory, and the computed bond lengths are presented in **Figure 4a**. Specifically, the optimized structures were determined for the following termination conditions: fully =O terminated $Cr_2TiC_2O_2$, fully –F terminated $Cr_2TiC_2F_2$, a mixed termination state $Cr_2TiC_2O_{1.33}F_{0.66}$, and non-terminated $Cr_2TiC_2$. Informed by these calculations, we assign the first EXELFS $|\chi(R)|$ peak to overlapping 1st nearest neighbor (NN) Cr-C and Cr-T bonds, and the second $|\chi(R)|$ peak to overlapping 2nd NN Cr-Cr and Cr-Ti bonds. With annealing, the most pronounced change in the $|\chi(R)|$ data is a decrease in the 1st NN peak (Figure 3b). As we detail below, this change is indicative of a decrease in the MXene surface termination concentration with annealing.

We performed quantitative data analysis by fitting the EXELFS data using the same procedures as in standard EXAFS data analysis.[34] Owing to the similar atomic number of O and F and the similar Cr-O and Cr-F bond lengths, all Cr-T bonds were represented with Cr-O in the fit. Likewise, all 2nd NN bonds were represented with Cr-Ti. Hence, the fitting model consisted of three bonds: Cr-C, Cr-T (represented with Cr-O), and Cr-2nd NN (represented with Cr-Ti). Each bond was fit with a bond length, $R$, and a bond length mean square deviation, $\sigma^2$. Assuming stoichiometry between Cr, C, and Ti, the Cr-C coordination number (CN) was fixed at 3 and the Cr-2nd NN CN was fixed at 9 (3 Cr-Ti bonds and 6 Cr-Cr bonds). The Cr-T CN was fixed at 3 for the AP spectra (assuming full termination), but allowed to vary for spectra collected after annealing (to simulate termination loss). With this strategy, we obtained an excellent fit to all spectra (Figure S4). The best fit bond lengths are given in Figure 4b and are in good agreement with the DFT values.

The main result from the EXELFS analysis is that the Cr-T CN substantially decreased with annealing. In keeping with the MXene literature, we plot the fitted number of termination species per $Cr_2TiC_2$ formula unit, which is the fitted EXELFS Cr-T CN multiplied by 2/3 (Figure 3c). For comparison, we also plot the concentration of F and O as determined through EELS composition analysis, with the aluminum fluoride signal subtracted from the F value (Figure 3c, Methods). Good agreement is observed between the total change in T determined with EXELFS fitting and the total change in the F and O composition determined with EELS. A clear picture of surface chemistry (and its thermal evolution) in $Cr_2TiC_2T_x$ is thus obtained through our combined approach. Initially, the ratio of =O to –F termination is roughly 2:1, giving a MXene chemistry of ~$Cr_2TiC_2O_{1.3}F_{0.8}$. Changes in surface functionalization are minimal with annealing up to 300 °C, but between 400 – 600 °C, all of the –F termination is lost, and the total number of termination species per $Cr_2TiC_2$ unit decreases by a factor of ~1/3. After annealing at 600 °C, the MXene chemistry is ~$Cr_2TiC_2O_{1.3}$ with 1/3 of the surface sites non-terminated. –OH termination may exist for the AP MXene, but if present, –OH is presumably transformed to =O under the electron beam.

An interesting trend in the DFT results – present across all levels of theory – is that the Cr-C and Cr-Ti bond lengths decrease going from surface terminated to non-terminated (Figure 4a). Experimentally, EXELFS analysis of the AP versus 600 °C annealed samples offer a comparison of fully terminated versus partially terminated MXene, where the predicted termination-lattice coupling may be present. However, changes in the fitted bond lengths with annealing show no clear trend (Figure 4b). Further improvement in the EXELFS resolution and signal-to-noise ratio are needed to perform such quantitative analysis in the future.[24,32]

Lastly, we comment on these results in relation to other MXenes. In terms of $Cr_2TiC_2T_x$'s thermal stability up to 700 – 800 °C in vacuum, $Cr_2TiC_2T_x$ is more stable than $Ti_2CT_x$, similar to $Ti_3CNT_x$ and $Ti_3C_2T_x$, and less stable than $Mo_2TiC_2T_x$.[14] Regarding the stability of the Cr-T bonds during vacuum annealing, the Cr-T bonds in $Cr_2TiC_2T_x$ appear weaker than the Ti-T bonds of Ti based MXenes, where similar in situ annealing treatments only lead to the partial loss of –F. The ability to thermally remove all –F terminations thus holds promise for termination engineering with $Cr_2TiC_2T_x$. The Cr-T bonds are, however, stronger than Mo-T bonds, as =O terminations can be removed from $Mo_2TiC_2T_x$ with annealing above 500 °C, but =O terminations are stable in $Cr_2TiC_2T_x$ up to at least 700 °C.[14]

## 3. Conclusion

In conclusion, we utilized in situ EELS with EXELFS analysis, synchrotron XAFS, and DFT calculations to investigate the surface functionalization and surface evolution of $Cr_2TiC_2T_x$ upon thermal annealing. We showed that vacuum annealing this MXene to 600 °C leads to the complete loss of –F termination, which could facilitate termination engineering of $Cr_2TiC_2T_x$ for magnetic and electronic property tuning. This work also showcases the promise of correlating conventional and high energy EELS and EXELFS for surface chemistry characterization.

## 4. Experimental Section/Methods

*MXene synthesis:* Parent $Cr_2TiAlC_2$ MAX powder was etched in a mixture of HF and HCl at 35 °C for 42 h. After etching, the reaction mixture was washed with deionized (DI) water via repeated centrifugation until the pH of the supernatant was close to ~7. To delaminate the MXene, it was added to tetramethylammonium hydroxide in water (~2.5 %v/v), stirred overnight at room temperature, and then washed in DI water and centrifuged. See ref. [19] for more details on the MXene synthesis and characterization.

*In situ TEM sample preparation and experiments:* All TEM and EELS measurements were performed on a JEOL 2100F TEM with a Gatan Imaging Filter Quantum and a K2 IS electron detector for EELS.[23] For the in situ EELS measurements (excluding the Cr *K*-edge experiment), the delaminated MXene solution was deposited via spray-casting onto a DENSsolutions heating and biasing nanochip. EELS measurements were performed in the same area after each annealing step. The microscope was operated in STEM mode, and the STEM probe rastered across an area of ~500 nm² during EELS acquisition. The EELS dispersion was 0.125 eV/channel, and the energy resolution was ~1 eV. Due to energy drift between measurements, all spectra were aligned to the Cr *L*-edge onset. Note that these MXene films were supported on a SiN underlayer, which makes determination of the local =O concentration difficult due to surface oxidation of the SiN. For all annealing steps, heating and cooling rates were set to 1 °C/s, and the maximum temperature was held for 5 mins.

For the in situ Cr *K*-edge EXELFS measurements, the MXene was drop-casted onto lacey carbon grids, and annealing was performed with the Gatan 626 hot stage. A large beam current of 7 nA was used to increase the signal-to-noise ratio at high energy, and a spread TEM beam was used to reduce the local current density. To minimize the likelihood of beam-induced damage, each new measurement was performed in a new area. For each annealing condition, two measurements (from separate areas) were collected and processed independently, and then merged prior to data fitting. Individual datasets are shown in Figure S5. For the 8 areas sampled, the thicknesses ranged from $t/\lambda = 0.7$ up to 1.1. A Fourier ratio deconvolution was performed to remove plural scattering, as implemented in GMS 3.31. The experimental dispersion was 0.5 eV/channel, but after calibration on Ni *K*-edge EXELFS data, the dispersion was set to 0.485 eV/channel during data processing.[24] For the Cr *K*-edge, the collection semi-angle was ~60 mRad. Owing to difficulties with high energy EELS calibration, all Cr *K*-edge EELS measurements were aligned in energy to the XAFS Cr *K*-edge measurement. Prior to each EXELFS measurement, low energy EELS data was collected for compositional analysis with a dispersion of 0.125 eV/channel and an energy resolution of ~1.5 eV. The atomic ratios between O, Cr, and F were determined using the O *K*-edge, Cr *L*-edge, and F *K*-edge using a power-law background subtraction, Hartree-Slater cross-sections, and accounting for plural scattering through ZLP measurements, as implemented in GMS 3.31. To determine –F, we subtracted the residual F concentration measured after the 600 °C anneal from all other datasets, thereby removing the Al fluoride contribution from the –F concentration.

*XAFS acquisition:* The Cr *K*-edge X-ray absorption spectra were collected at 7-BM (QAS) beamline, National Synchrotron Light Source II, Brookhaven National Laboratory (BNL). Samples were prepared by vacuum filtration of MXene solution, forming free-standing MXene film. The data were collected in transmission mode.

*EXELFS and EXAFS data processing:* Cr *K*-edge EELS and XAFS data were normalized using the Athena software package.[34] Low order polynomials were fit to the pre- and post-edge regions, allowing edge step normalization of the absorption coefficient. All EXELFS fitting was performed with the Artemis software package. For fitting, the data was $k^2$ weighted, the *k*-range was $2 - 9$ Å$^{-1}$, and the *R*-range was $1 - 2.9$ Å.

*Density functional theory calculations:* Density functional theory (DFT) calculations were carried out with the Vienna ab initio simulation package VASP 5.4.[35,36] The Kohn-Sham equations are solved using a plane-wave basis set, applying the projected augmented wave (PAW) potential.[36] The plane-wave cutoff energy was set to 500 eV, tested to ensure the convergence of total energy is within 0.01 eV/atom. Cut-off energies of 550 eV and 600 eV, as previously suggested[21,37] resulted in changes in the total energy of 0.001-0.003 eV. Structure optimizations, including the lattice constants and atomic internal positions, were performed with the generalized gradient approximation (GGA) PBE functional,[38] and for comparison, also with PBE+U (U=4.0 eV was previously applied),[21] including Grimme's correction for London dispersion (PBE+U+D3),[39] as well as with the range-separated hybrid HSE-06 exchange-correlation functional.[40,41] A 3×2 cell was used, enabling modeling of the mixed termination (33% F and 66% O), such that each surface layer has 6 atoms of the same atomic species except for the mixed layer where there are 2 F atoms and 4 O atoms. Spin-polarization was also considered. Distances are given as averages of the possible distances for different atomic layers. A 3×4×1 Monkhorst-Pack *k*-point grid was used, tested to ensure total energy convergence within 0.01 eV, with convergence criteria set so that the difference in the total energy is within $1\times10^{-8}$ eV per cell, and forces were less than 0.01 eV/Å. A vacuum region of 14 Å - 15 Å was used, depending on the termination type.

**Acknowledgements**
J.L.H., A.C.L, and M.L.T. acknowledge funding from the National Science Foundation (NSF) MRI award #DMR-1429661. Y. L. and A.I.F. acknowledge support from the NSF award #DMR-1911592 for the XAS measurements and XAS and EXELFS data analyses. K.H. and Y.G. acknowledged financial support from the U.S. Department of Energy (DOE), Office of Science, Office of Basic Energy Sciences, grant no. DE-SC0018618. The Cr *K*-edge XAS data were acquired at beamline 7-BM (QAS) of the National Synchrotron Light Source II (NSLS-II), a U.S. DOE Office of Science User Facility operated for the DOE Office of Science by Brookhaven National Laboratory (BNL) under Contract No. DE- SC0012704. Prof. Babak Anasori is acknowledged for providing $Cr_2TiAlC_2$ MAX phase used in this study. The authors declare no conflicts of interest.

**References**
[1] F. Shahzad, M. Alhabeb, C. B. Hatter, B. Anasori, S. M. Hong, C. M. Koo, Y. Gogotsi, *Science.* **2016**, *353*, 1137.
[2] A. Sarycheva, A. Polemi, Y. Liu, K. Dandekar, B. Anasori, Y. Gogotsi, *Sci. Adv.* **2018**, *4*, eaau0920.
[3] N. K. Chaudhari, H. Jin, B. Kim, D. San Baek, S. H. Joo, K. Lee, *J. Mater. Chem. A* **2017**, *5*, 24564.
[4] M. Naguib, M. Kurtoglu, V. Presser, J. Lu, J. Niu, M. Heon, L. Hultman, Y. Gogotsi, M. W. Barsoum, *Adv. Mater.* **2011**, *23*, 4248.
[5] K. Hantanasirisakul, Y. Gogotsi, *Adv. Mater.* **2018**, *30*, 1804779.
[6] L. Verger, V. Natu, M. Carey, M. W. Barsoum, *Trends Chem.* **2019**, *1*, 656.
[7] M. Khazaei, A. Ranjbar, M. Arai, T. Sasaki, S. Yunoki, *J. Mater. Chem. C* **2017**, *5*, 2488.
[8] J. Zhu, E. Ha, G. Zhao, Y. Zhou, D. Huang, G. Yue, L. Hu, N. Sun, Y. Wang, L. Y. S. Lee, C. Xu, K. Y. Wong, D. Astruc, P. Zhao, *Coord. Chem. Rev.* **2017**, *352*, 306.
[9] M. Alhabeb, K. Maleski, B. Anasori, P. Lelyukh, L. Clark, S. Sin, Y. Gogotsi, *Chem. Mater.* **2017**, *29*, 7633.
[10] V. Kamysbayev, A. Filatov, H. Hu, X. Rui, F. Lagunas, D. Wang, R. Klie, D. Talapin, *Science.* **2020**, *8311*, eaba8311.
[11] C. Wang, S. Chen, L. Song, *Adv. Funct. Mater.* **2020**, *2000869*, 1.
[12] T. Schultz, N. C. Frey, K. Hantanasirisakul, S. Park, S. J. May, V. B. Shenoy, Y. Gogotsi, N. Koch, *Chem. Mater.* **2019**, *31*, 6590.
[13] M. Anayee, N. Kurra, M. Alhabeb, M. Seredych, M. N. Hedhili, A. H. Emwas, H. N. Alshareef, B. Anasori, Y. Gogotsi, *Chem. Commun.* **2020**, *56*, 6090.
[14] J. L. Hart, K. Hantanasirisakul, A. C. Lang, B. Anasori, D. Pinto, Y. Pivak, J. T. Van

Omme, S. J. May, Y. Gogotsi, M. L. Taheri, *Nat. Commun.* **2019**, *10:522*.

[15] M. A. Hope, A. C. Forse, K. J. Griffith, M. R. Lukatskaya, M. Ghidiu, Y. Gogotsi, C. P. Grey, *Physcal Chem. Chem. Phys.* **2016**, *18*, 5099.

[16] I. Persson, L.-A. Naslund, J. Halim, M. W. Barsoum, V. Darakchieva, J. Palisaitis, J. Rosen, P. O. A. Persson, *2D Mater.* **2018**, *5*, 015002.

[17] H. W. Wang, M. Naguib, K. Page, D. J. Wesolowski, Y. Gogotsi, *Chem. Mater.* **2016**, *28*, 349.

[18] J. Halim, K. M. Cook, M. Naguib, P. Eklund, Y. Gogotsi, J. Rosen, M. W. Barsoum, *Appl. Surf. Sci.* **2016**, *362*, 406.

[19] K. Hantanasirisakul, B. Anasori, S. Nemsak, J. L. Hart, J. Wu, Y. Yang, R. V Chopdekar, P. Shafer, A. F. May, E. J. Moon, J. Zhou, Q. Zhang, M. L. Taheri, S. J. May, Y. Gogotsi, *Nanoscale Horizons* **2020**, *5*, 1557.

[20] J. Yang, X. Zhou, X. Luo, S. Zhang, L. Chen, *Appl. Phys. Lett.* **2016**, *109*, 203109.

[21] W. Sun, Y. Xie, P. R. C. Kent, *Nanoscale* **2018**, *10*, 11962.

[22] J. Zhou, X. H. Zha, M. Yildizhan, P. Eklund, J. Xue, M. Liao, P. O. Å. Persson, S. Du, Q. Huang, *ACS Nano* **2019**, *13*, 1195.

[23] J. L. Hart, A. C. Lang, A. C. Leff, P. Longo, C. Trevor, R. D. Twesten, M. L. Taheri, *Sci. Rep.* **2017**, *7*, 8243.

[24] J. L. Hart, A. C. Lang, Y. Li, K. Hantanasirisakul, A. I. Frenkel, M. L. Taheri, *ArXiv Prepr. 1909.06323* **2019**.

[25] D. Magne, V. Mauchamp, S. Celerier, P. Chartier, T. Cabioc'h, *Physcal Chem. Chem. Phys.* **2016**, *18*, 30946.

[26] A. Feng, Y. Yu, Y. Wang, F. Jiang, L. Mi, L. Song, *Mater. Des.* **2017**, *114*, 161.

[27] M. Seredych, C. E. Shuck, D. Pinto, M. Alhabeb, E. Precetti, G. Deysher, B. Anasori, N. Kurra, Y. Gogotsi, *Chem. Mater.* **2019**, *31*, 3324.

[28] M. Ashton, K. Mathew, R. G. Hennig, S. B. Sinnott, *J. Phys. Chem. C* **2016**, *120*, 3550.

[29] T. Hu, Z. Li, M. Hu, J. Wang, Q. Hu, Q. Li, X. Wang, *J. Chem. Phys.* **2017**, *121*, 19254.

[30] H. Zhang, T. Hu, W. Sun, M. Hu, R. Cheng, X. Wang, *Chem. Mater.* **2019**, *31*, 4385.

[31] Y. Xie, M. Naguib, V. N. Mochalin, M. W. Barsoum, Y. Gogotsi, X. Yu, K. Nam, X. Yang, A. I. Kolesnikov, P. R. C. Kent, *J. Am. Chem. Soc.* **2014**, *136*, 6385.

[32] I. Maclaren, K. J. Annand, C. Black, A. J. Craven, *Microscopy* **2018**, *67*, 78.

[33] A. J. Berry, H. S. C. O'Neill, *Am. Mineral.* **2004**, *89*, 790.

[34] B. Ravel, M. Newville, *J. Synchrotron Radiat.* **2005**, *12*, 537.

[35] G. Kresse, J. Furthmüller, *Comput. Mater. Sci.* **1996**, *6*, 15.

[36] G. Kresse, D. Joubert, *Phys. Rev. B - Condens. Matter Mater. Phys.* **1999**, *59*, 1758.

[37] D. Sun, Q. Hu, J. Chen, X. Zhang, L. Wang, Q. Wu, A. Zhou, *ACS Appl. Mater. Interfaces* **2016**, *8*, 74.

[38] J. P. Perdew, K. Burke, M. Ernzerhof, *Phys. Rev. Lett.* **1996**, *77*, 3865.

[39] S. Grimme, J. Antony, S. Ehrlich, H. Krieg, *J. Chem. Phys.* **2010**, *132*, 154104.

[40] A. V. Krukau, O. A. Vydrov, A. F. Izmaylov, G. E. Scuseria, *J. Chem. Phys.* **2006**, *125*.

[41] J. Heyd, G. E. Scuseria, M. Ernzerhof, *J. Chem. Phys.* **2003**, *118*, 8207.

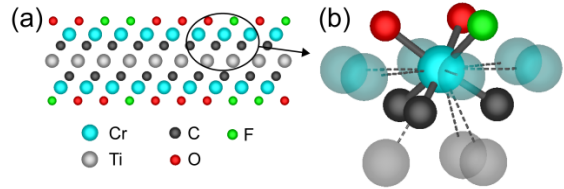

**Figure 1.** (a) Schematic view along the *a* axis of the $Cr_2TiC_2T_x$ structure with mixed =O and –F termination. (b) Magnified view of the local structure of each surface Cr atom. $1^{st}$ nearest neighbor bonds are shown with thick, solid lines. $2^{nd}$ nearest neighbor bonds are shown with thin, dashed lines, and translucent atoms. Surface terminations are assumed follow FCC stacking relative to the underlying C and Cr layers.

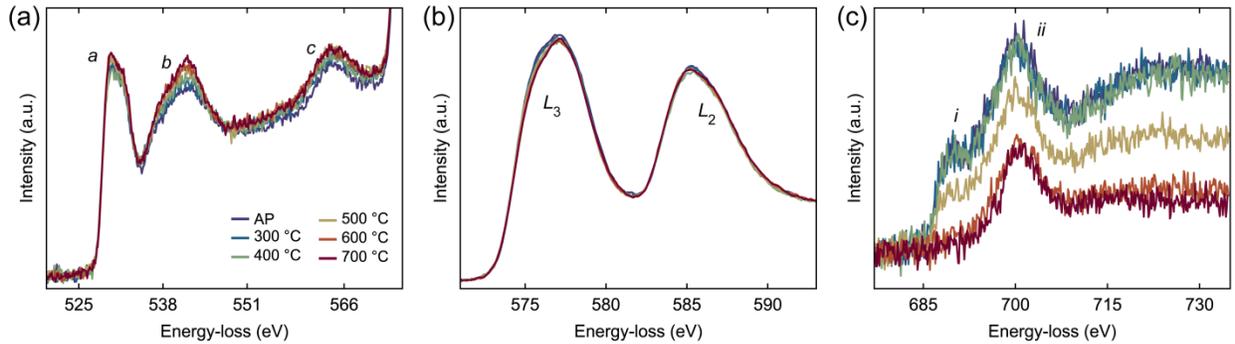

**Figure 2.** In situ EELS measurements of the O *K*-edge (a), Cr *L*-edge (b), and F *K*-edge (c). The legend in (a) applies to (b) and (c) as well. All spectra were collected at room temperature after annealing. The O *K*-edge and F *K*-edge were normalized to the Cr *L*-edge intensity. Cr *L*-edge spectra are normalized to their post-edge continuum. Ti *L*-edge measurements are shown in Figure S2.

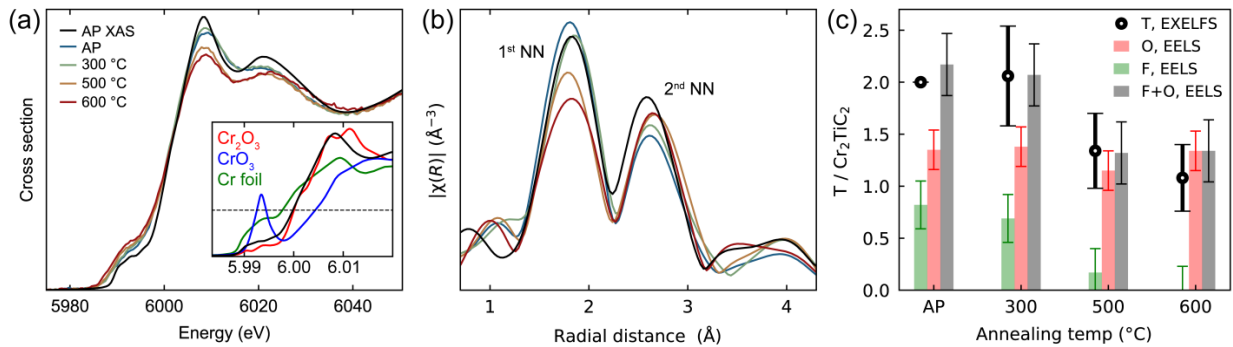

**Figure 3.** (a) Normalized EELS and XAFS Cr $K$-edge measurements of $Cr_2TiC_2T_x$ in the near-edge region. All data collected at room temperature after annealing. The inset compares the XAFS measurement of as-prepared (AP) MXene (in black) alongside XAFS reference spectra, and the x-tick marks have units of keV. The horizontal dashed line is at 0.5 normalized intensity, and helps visualize the trend in edge onset with oxidation state. (b) Fourier transform magnitudes ($|\chi(R)|$, $k^2$-weighted, corrected for phase shift) of the Cr $K$-edge data displayed in (a). The Fourier transform range was $2 - 9$ Å$^{-1}$. The $\chi(k)$ data is shown in Figure S3. The legend in (a) applies to (b) as well. (c) The total number of termination species per $Cr_2TiC_2$ formula unit as determined from EXELFS fitting (circular markers) along with the F and O concentrations as determined from EELS composition analysis (bars). Note that the F and O concentrations were not determined from the spectra shown in Figure 2, rather, additional O and F $K$-edges were acquired during the Cr $K$-edge measurements, which was performed on a separate sample.

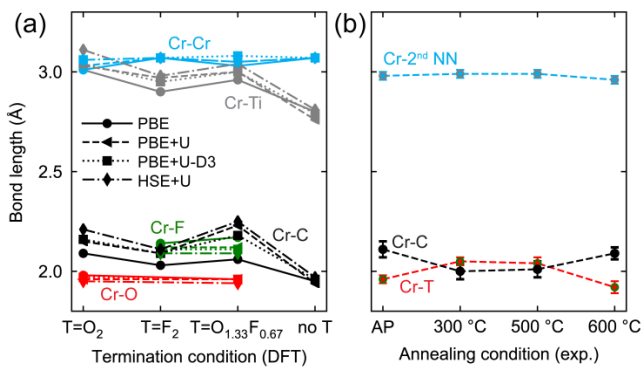

**Figure 4.** (a) Comparison of the DFT computed bond lengths at various levels of theory. Each bond type is shown with a different color, and each DFT method is shown with a different marker and line type (see middle left legend). (b) Fitted bond lengths from EXELFS analysis as a function of in situ annealing. The Cr-T bond represents both Cr-O and Cr-F bonds, and the Cr-2$^{nd}$ NN bond represents both Cr-Cr and Cr-Ti bonds. The y-axis of (b) is the same as (a).

11